\documentclass[12pt]{article}

\usepackage{latexsym}
\usepackage{amsmath}
\usepackage[margin=1in,lines=25]{geometry}
\usepackage{amsfonts}
\usepackage{amssymb}
\usepackage{setspace}
\usepackage{psfrag}
\usepackage{graphicx,bm,physics}
\renewcommand{\hat}{\widehat}
\renewcommand{\bar}{\overline}
\newcommand{\pr}{\mbox{pr}}

\newtheorem{prop}{Proposition}
\usepackage{booktabs}
\usepackage{caption}
\usepackage{natbib}
\usepackage[colorlinks, citecolor=blue]{hyperref}

\renewcommand{\eqref}[1]{(\ref{#1})}

\title{Zero Inflated Poisson
Model with Clustered
Regression Coefficients: an Application to Heterogeneity Learning
of Field Goal Attempts of  Professional Basketball Players}
\author{Guanyu Hu ~~~ Hou-Cheng Yang ~~~ Yishu Xue ~~~ Dipak K. Dey}
\begin{document}
\maketitle 
\begin{abstract}
    Although basketball is a dynamic process sport, with 5 plus 5 players competing
on both offense and defense simultaneously, learning some static information is
predominant for professional players, coaches and team mangers. In order to have
a deep understanding of field goal attempts among different players, we propose
a zero inflated Poisson model with clustered regression coefficients to learn
the shooting habits of different players over the court and the heterogeneity
among them. Specifically, the zero inflated model recovers the large proportion
of the court with zero field goal attempts, and the mixture of finite mixtures
model learn the heterogeneity among different players based on clustered
regression coefficients and inflated probabilities. Both theoretical and
empirical justification through simulation studies validate our proposed method.
We apply our proposed model to the National Basketball Association (NBA), for
learning players' shooting habits and heterogeneity among different players over
the 2017--2018 regular season. This illustrates our model as a way of providing
insights from different aspects.\bigskip \\
keywords: 
Bayesian Nonparametric; MCMC; Mixture of Finite Mixtures; Model Based
Clustering
\end{abstract}

\section{INTRODUCTION}

Analyzing players' ``hotspots'', i.e., locations where they make the most
shooting attempts, is an indispensable part of basketball data
analytics. Identifying such hotspots, as well as which players tend to have
similar hotspot locations, provides valuable information for coaches as well as
for teams who are aiming at making transactions and looking for players of a
specific type. One preliminary tool for representing shot locations is the shot
chart, but it is rather rough as there is no clear-cut way of defining
``similarity'', which calls for the need for more rigorous statistical
modeling.

Various tools have been proposed to model point patterns. Among them,
spatial point processes is a family of models that assume event locations are
random, and realized from an underlying process, which has an intensity
surface. Spatial point processes have a wide range of variants, the most
prominent of which being the Poisson process \citep{geyer1998likelihood}, the
Gibbs process \citep{goulard1996parameter}, and the log-Gaussian Cox process
\citep[LGCP;][]{moller1998log}. They also have a wide range of applications,
including ecological studies, environmental sciences
\citep{jiao2020heterogeneity,hu2019new}, and sports analytics.
\cite{reich2006spatial} developed a multinomial logit model that incorporates
spatially varying coefficients, which were assumed to follow a heterogeneous
Poisson process.
\cite{miller2014factorized} discussed creating low-dimensional representation
of players' shooting habits using several different spatial point processes.
These works, however, focus mainly on characterizing the shooting behavior of
individual players. Which players are similar to each other, however, remains
un-answered by these works.

Towards this end, \cite{jiao2019bayesian} proposed a marked point process joint
modeling approach that takes into account both shot locations and outcomes. The
fitted model parameters are grouped using \emph{ad hoc} approaches to identify
similarities among players.
In a study of tree
locations, \cite{jiao2020heterogeneity} proposed a model-based clustering
approach that incorporates the Chinese restaurant process
\citep[CRP;][]{ferguson1973bayesian} to account for the latent grouped
structure. The number of clusters is readily inferred from the number of unique
latent cluster labels. \cite{yin2020bayesian} improved the model of
\cite{jiao2020heterogeneity} by using Markov random fields constraint Dirichlet
process for the latent cluster belongings, which effectively encourages local
spatial homogeneity. \cite{hu2020bayesiangroup} used LGCP to obtain the
underlying intensity, and then defined a similarity measure on the intensities
of different players, which was later used in a hierarchical model that
employed mixture of finite mixtures
\citep[MFM;][]{miller2018mixture} to perform clustering. \cite{yin2020analysis}
proposed a Bayesian nonparametric matrix clustering approach to analyze the
latent heterogeneity structure of estimated intensity surfaces.
Note that in all five works, the intensity function always played a certain
role, which adds another layer of modeling between the shots and the grouping
structure.

One natural way to model the counts directly without employing the intensity
surface is the Poisson regression. \cite{zhao2020bayesian} proposed a spatial
homogeneity pursuit regression model for count value data, where clustering of
locations is done via imposing certain spatial contiguity constraints on MFM.
Data of basketball shots, however, poses more challenges. The first challenge
comes from the fact that only few shots are made by players in the region
near the half court line, which means there is a large portion of the court
that corresponds to no attempts. Secondly, existing approaches either only
perform clustering on the spatial domain
\citep{zhao2020bayesian,yin2020bayesian}, or utilize the intensity surface and
cluster players in terms of shooting habits \citep{hu2020bayesiangroup}.
Thirdly, to demonstrate its superiority over heuristic comparison and grouping,
a model-based approach should have favorable theoretical properties such as
consistent estimation for the number of clusters and clustering configurations.

To tackle the three challenges, we propose a Bayesian zero-inflated Poisson
(ZIP) regression approach to model field goal attempts of players with
different shooting habits. The contribution of this paper is three-fold. First,
the large proportion of the court with zero shot attempts is accommodated in
the model structure by zero inflation. Next, non-negative matrix factorization
is utilized to decompose the shooting habits of players into linear
combinations of several basis functions, which naturally handles the homogeneity
pursuit on the spatial dimension. On the dimension across players, we for the
first time introduce a MFM
prior in ZIP model to jointly estimate regression coefficients and zero
inflated probability and their clustering information.
Finally, we provide both theoretical and empirical justification through
simulations for the model's performance in terms of both estimation and
clustering.

The rest of the paper is organized as follows. In Section
\ref{sec:motivatedata}, we introduce the motivating data from 2017--2018
regular season. In Section \ref{sec:method}, we first review zero inflated
Poisson regression, and then propose our Bayesian clustering method based on MFM.
Details of Bayesian inference are presented in Section
\ref{sec:bayesian_inference}, including the MCMC algorithm and post MCMC
inference methods. Simulation studies are conducted in Section
\ref{sec:simulation}. Applications of the proposed methods to NBA players data
are reported in Section \ref{sec:realdata}. Section \ref{sec:discussion}
concludes the paper with a discussion.

\section{Motivating Data}\label{sec:motivatedata}

Our data consists of both made and missed field goal attempt locations from the
offensive half court of games in the 2017--2018 National Basketball Association
(NBA) regular season. The data is available
at~\url{http://nbasavant.com/index.php}, and also on GitHub
(\url{https://github.com/ys-xue/MFM-ZIP-Basketball-Supplemental}). 
We focus on
players that have made more than 400 field goal attempts. Also, players who just
started their careers in the 2017--2018 season, such as Lonzo Ball and Jayson
Tatum, are not considered. A total of 191 players who meet the two criteria
above are included in our analysis.

\begin{figure}[t]
	\centering
	\includegraphics[width=\textwidth]{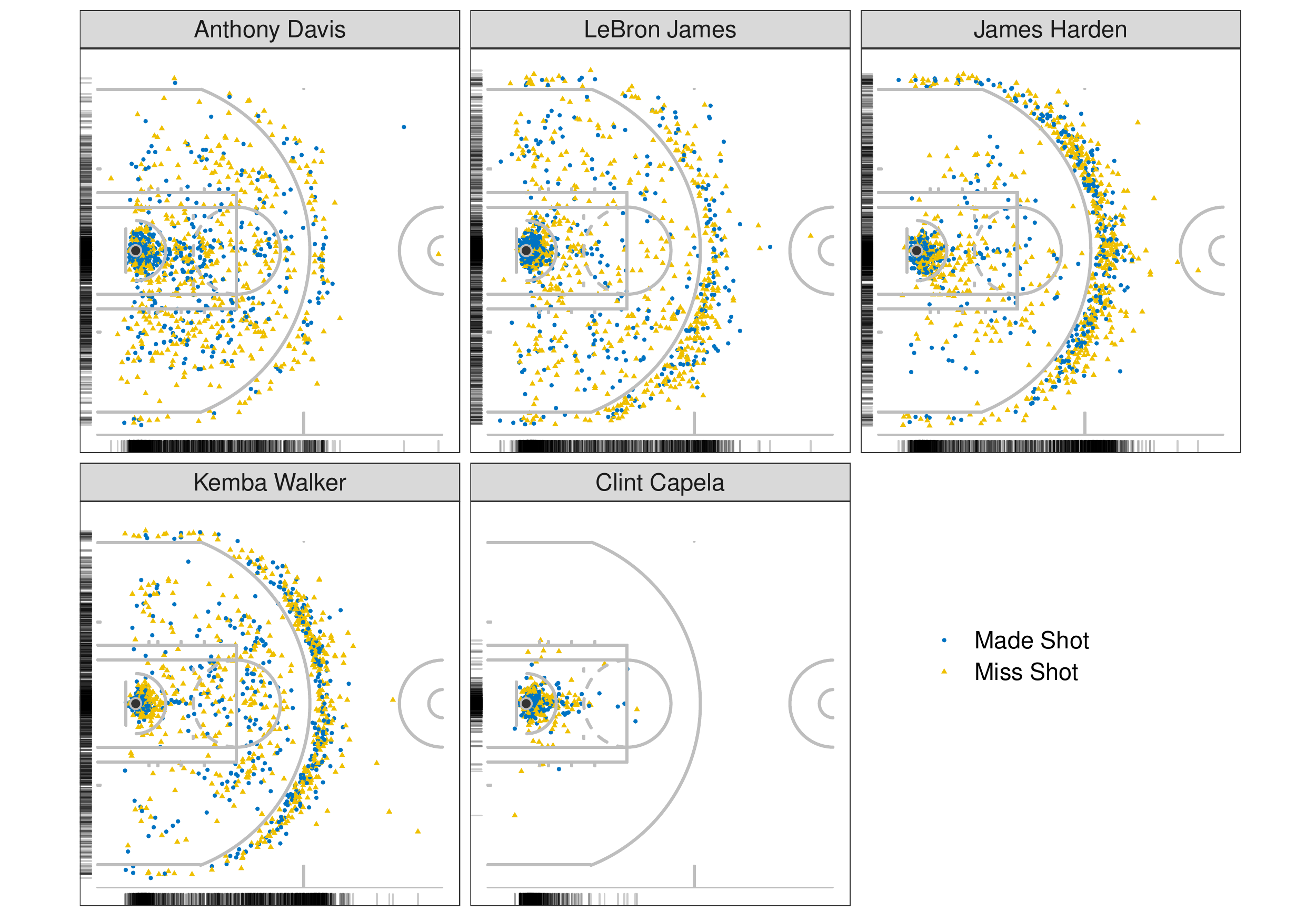}
	\caption{\label{fig:shotchartNMF} Shot charts for selected NBA players.}
\end{figure}

We model a player’s shooting location choices and outcomes as a spatial point
pattern on the offensive half court, a 47 ft by 50 ft rectangle, which is the
standard size for NBA. The spatial domain for the basketball court is denoted
as
$D\in\left[0, 47\right] \times \left[0, 50\right]$. We partition the court to
1~ft $\times$ 2~ft blocks, which means that there are in total $47\times25=1175$
blocks in the basketball court. The shot charts for five selected
players are visualized in Figure~\ref{fig:shotchartNMF}.
The numbers of shot attempts in each of the blocks are counted.
Hence, this data consists of non-negative, highly skewed sequence
counts with a large proportion of zeros, as most shots are made in the range
from the painted area to the three-point line, and many of the blocks between
the three-point line and mid-court line have no corresponding positive values.
This abundance of 0's motivates the usage of zero-inflated models for such
type of data. Here we define
$\bm{y}=\left(\bm{y}_1,\bm{y}_2,\ldots,\bm{y}_n\right)$
where $\bm{y}_i=(y_{i1},y_{i2},\ldots,y_{iJ})^\top$ for $i=1,2,\ldots,191$ and $J = 1175$.
Each $y_{ij}$ for $i=1,\ldots, 191$ and $j=1,\ldots, J$
represents the total number of shots made by
the $i$-th player in the $j$-th block. For selected players, we counted the
number of blocks that have no shot, one shot, two shots, $\ldots$, and more
than six shots as presented in Table~\ref{tab:data}. One thing that can be
noticed that Clint Capela has the most number of blocks corresponding to zero
shots, which is straightforward as he is center, and barely shoots out of the
painted area. LeBron James, on the other hand, has the least number of blocks
with no shots, which indicates his wide shooting range. Except for Capela,
the four other players have non-trivial, positive number of blocks
corresponding to 1, 2, 3, 4, 5, and 6+ shots, indicating that they are
comfortable shooting from a larger range. James Harden in particular, has 45
blocks with 6+ shots, indicating that he has the largest range of ``hotspots''.

\begin{table}
\centering 
\caption{\label{tab:data} Number of blocks corresponding to number of shots
for selected players.}
	\begin{tabular}{cccccc}
\hline
Observed Value & \text{Davis} &\text{James}&
		\text{Harden}&\text{Walker}&\text{Capela}\\
\hline
		0 & 836 & 788 & 853 & 837 & 1123 \\
		1 & 161 & 212 & 149 & 159 & 25\\
		2 & 88 & 91 & 69 & 82 & 8 \\
		3 & 38 & 35 & 31 & 35 & 2\\
		4 & 18 & 16 & 23 & 17 & 4\\
		5 & 9 & 8 & 5 & 9 & 0\\
		6+& 25 & 25 & 45 & 36 & 13 \\
\hline
	\end{tabular}
\end{table}

\section{Methodology}\label{sec:method}
\subsection{Zero Inflated Poisson Regression}

In this section, we briefly discuss the zero-inflated Poisson distribution
(ZIP). There are some models are capable of dealing with excess zero counts
\citep{mullahy1986specification,lambert1992zero}, and zero-inflated models are one
type of them. Zero-inflated models are two-component mixture models that
combine a count component and a point mass at zero with a count distribution
such as Poisson, geometric or negative binomial \citep[see,][for a
discussion]{cameron2005microeconometrics}. We denote the observed count values
of the $i$-th player as
$\bm{y}_i=\left(y_{i1},y_{i2},\ldots,y_{i,1175}\right)^\top$. Hence, the
probability distribution of the ZIP random variable $y_{ij}$ can be written as,
for a nonnegative integer~$\kappa$, 
\begin{align}
    \pr(y_{ij}=\kappa)=
    \begin{cases}
      \rho_i+\left(1-\rho_i\right)\exp{-\mu_{ij}} & \text{if $\kappa=0$}\\
	  \left(1-\rho_i\right)\frac{\mu_{ij}^{y_{ij}}\exp{-\mu_{ij}}}{y_{ij}!} & 
	  \text{if $\kappa>0$}
    \end{cases},  
    \label{zip}
\end{align}
where~$\rho_i$ is the probability of extra zeros and~$\mu_{ij}$ is the mean
parameter of Poisson distribution.
For the rest of the paper, we denote this distribution as~$\mbox{ZIP}(\mu_{ij}, \rho_i)$.
It can be seen from \eqref{zip} that ZIP
reduces to the standard Poisson model when~$\rho_i=0$. Also, we know that
$\pr(y_{ij}=0)>\exp{-\mu_{ij}}$ indicates zero-inflation. The mean parameter
$\mu_{ij}$ is linked to the explanatory variables through log links as
\begin{align*}
    \log\left(\mu_{ij}\right)=\bm{x}_j^\top\bm{\beta}_i,
\end{align*}
where~$\bm{x}_j$ is a vector of covariates $\bm{x}_j=\left(
1, x_{1,j},\ldots, x_{p,j}\right)^\top$ and
$\bm{\beta}_i=\left(\beta_{0i},\beta_{1i},\ldots,\beta_{pi}\right)^\top$ are
the
corresponding regression coefficients including the intercept~$\beta_{0i}$.

\subsection{Non-Negative Matrix Factorization for Spatial Basis}\label{sec:NMF}

To capture shot styles of individual players, following
\cite{jiao2019bayesian},
we construct spatial basis functions using historical data. Shot data for a
total of~359 players who made over~100 shots in regular season 2016--2017 is
used as input. First, kernel density estimation is employed to estimate the
shooting frequency matrix~$\bm{\lambda}= (\lambda_1,\ldots, \lambda_{1175})$ for each
individual player. Similar checking of empirical correlation between the kernel
density values on blocks on the court is performed as in
\cite{miller2014factorized}, and the existence of long-range correlations in
non-stationary patterns motivates the usage of a basis construction method that
captures such long range correlation via global spatial patterns. This need
motivates the usage of non-negative matrix factorization
\citep[NMF;][]{sra2005generalized}
in our modeling effort.

NMF is a dimensionality reduction technique that assumes a matrix
$\bm{\Lambda}$
can be approximated by the product of two low-rank matrices 
\begin{align*}
    \bm{\Lambda}\approx \bm{W}\bm{B},
\end{align*}
where the matrix $\bm{\Lambda}\in\mathbb{R}_{+}^{N\times V}$ is composed of~$N$
data points of length~$V$, the basis matrix~$\bm{B}\in\mathbb{R}_{+}^{K\times
V}$ is composed of~$K$ basis vectors, and the weight matrix~$\bm{W}\in
\mathbb{R}_{+}^{N\times K}$ is composed of the~$N$ non-negative weight vectors
that scale and linearly combine the basis vectors to
reconstruct~$\bm{\Lambda}$.
The matrices~$\bm{W}$ and~$\bm{B}$ are obtained by minimizing certain
divergence
criteria (e.g. Kullback--Leibler divergence, or Euclidean distance), with the
constraint that all matrix elements remain non-negative. With the
non-negativity
restriction for both the weight vectors and basis vectors, NMF eliminates
redundant cases where negative bases ``cancel out'' positive bases. The basis
left is often more sparse, and focuses on partial representations, which can be
combined to represent the whole story \citep{lee1999learning}.

\begin{figure}
	\centering
	\includegraphics[width=\textwidth]{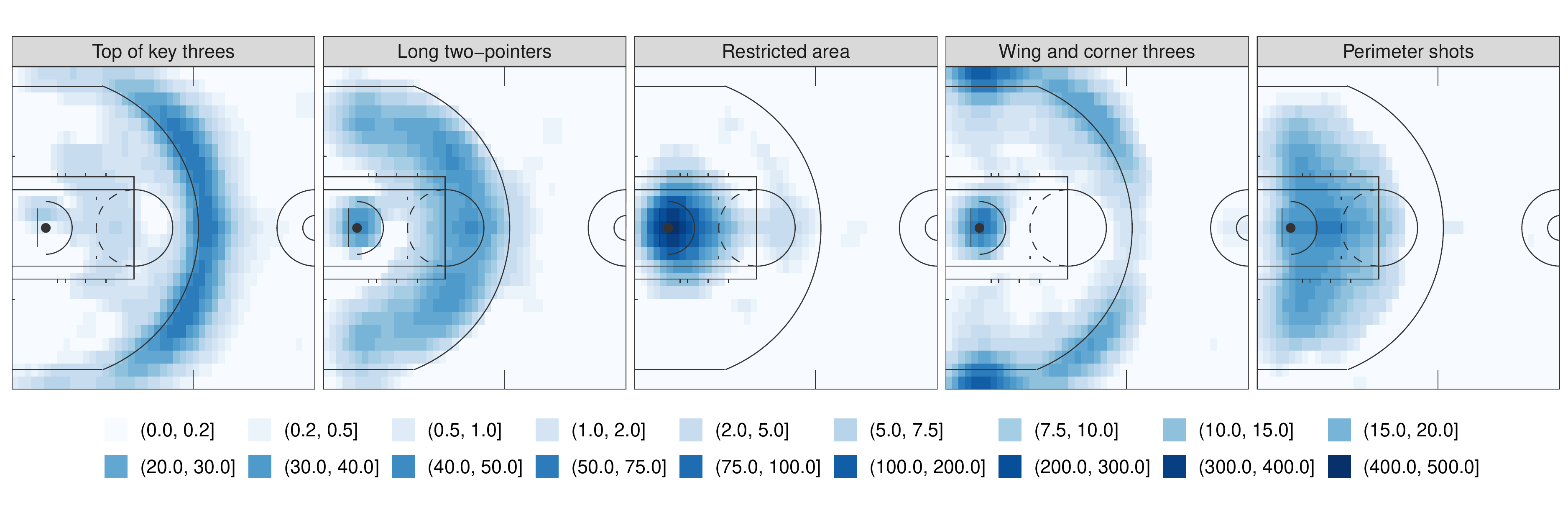}
	\caption{Visualization of basis functions obtained by NMF for $K = 5$. Each
	basis function represents the intensity function of a particular shot type.}
	\label{fig:basis5}
\end{figure}

In our application, with data for the 359 players as a~$359\times 1175$ matrix,
we use the \textsf{R} package \textbf{NMF} \citep{gaujoux2010flexible} to
obtain
$K=5$ bases, each of which correspond to a certain shot type, as illustrated in
Figure~\ref{fig:basis5}. These five bases correspond to, respectively, top of
key threes, long two-pointers, restricted area, wing and corner threes and
perimeter shots. A player's shooting habit can be approximated by a weighted
combination of these bases. Nevertheless, looking at the scales, one can see
that the maximum value is over~400, and the values are highly skewed. Natural
logarithm of the basis values is taken to reduce extreme values, and a
normalization step is subsequently performed so that values in each basis
vector
have mean~0 and standard deviation~1. When such normalized basis functions are
used as covariates in modeling individual player's shots, their corresponding
coefficients, or weight vectors, can be regarded as a characterization for the
shooting style of a player.

\subsection{ZIP with Clustered Regression Coefficients}
Consider a zero inflated Poisson regression model with varying coefficients as
follows,
\begin{align}
&y_{ij}\sim\text{ZIP}\left(\exp(\bm{x}_j^\top\bm{\beta}_i),\rho_i\right),
\quad i=1,\ldots, n, \quad j = 1,\ldots,J.
\label{eq:ZIP_model}
\end{align}
where~$\bm{\beta}_i$ is a~$p+1$-dimensional regression coefficients. From
\cite{gelfand2003spatial}, a Gaussian process prior can be assigned on
regression coefficients to obtain varying patterns. Assuming the existence of
clusters in $\bm{\beta}$ and~$\rho$, and denote the true cluster label for
player~$i$ as~$z_i$, the parameters  for cluster~$i$ as~$\bm{\beta}_{z_i}$
and~$\rho_{z_i}$, then the model in \eqref{eq:ZIP_model} can be rewritten as 
\begin{align}
&y_{ij}\sim\text{ZIP}\left(\exp(\bm{x}_j^\top\bm{\beta}_{z_i}),\rho_{z_i
}\right),
\quad i=1,\ldots, n,\quad j = 1,\ldots,J.
\end{align}
where $z_i\in\left\{1,\ldots, k\right\}$, with~$k$ being the total number of
clusters.

One popular way to model the joint distribution of $z_1,\ldots,z_k$ is the
Chinese restaurant process \citep[CRP;][]{blackwell1973ferguson}, which
defines a series of conditional distributions, also know as the a P\'{o}lya urn
scheme, as:
	\begin{eqnarray}\label{eq:crp}
	P(z_{i} = c \mid z_{1}, \ldots, z_{i-1})  \propto   
	\begin{cases}
	\abs{c}  , &  \text{at an existing table labeled}\, c\\
	\alpha,  & \text{if} \, $c$\,\text{is a new table}
	\end{cases},
	\end{eqnarray}
where $\abs{c}$ is the number of elements in cluster $c$.
Despite its favorable property as a method to simultaneously estimate the
number of clusters and the clustering configuration, it has proved to produce
extraneous clusters in the posterior, even when the number of sample size goes
to infinity, which renders the estimation for number of clusters inconsistent
\citep{miller2018mixture}. A slowed-down version of the CRP in terms of
producing new tables, mixture of finite mixtures
\citep[MFM;][]{miller2018mixture} is proposed to mitigate this problem:
\begin{eqnarray}\label{eq:MFM}
	k \sim p(\cdot), \quad (\pi_1, \ldots, \pi_k) \mid k \sim \mbox{Dir} (\gamma,
	\ldots, \gamma), \quad z_i \mid k, \pi \sim \sum_{h=1}^k  \pi_h \delta_h,\quad
	i=1, \ldots, n, 
\end{eqnarray}
where~$p(\cdot)$ is a proper probability mass function on~$\{1, 2, \ldots, \}$
and~$\delta_h$ is a point-mass at~$h$. Compared to the CRP, the introduction of
new tables is slowed down by the factor~$V_n(t+1)/ V_n(t)$, which allows for a
model-based pruning of the tiny extraneous clusters.
The coefficient~$V_n(t)$ is precomputed as
\begin{align*} 
	\begin{split}
	V_n(t) &= \sum_{n=1}^{+\infty}\dfrac{k_{(t)}}{(\gamma k)^{(n)}} p(k),
	\end{split}         
\end{align*} 
where $k_{(t)}=k(k-1)\ldots(k-t+1)$, and $(\gamma k)^{(n)} = {\gamma k} (\gamma
k+1)\ldots(\gamma k+n-1)$. The conditional distributions of $z_i, i=2, \ldots,
n$ under~\eqref{eq:MFM} can be defined in a P\'{o}lya urn scheme similar to
CRP:
\begin{eqnarray}\label{eq:mcrp}
	P(z_{i} = c \mid z_{1}, \ldots, z_{i-1})  \propto   
	\begin{cases}
	\abs{c} + \gamma  , &  \text{at an existing table labeled}\, c\\
	V_n(t+1)/ V_n(t)\gamma,  &  \text{if} \, c\,
	\text{is a new table}
	\end{cases},
\end{eqnarray}
with~$t$ being the number of existing clusters. Adapting MFM to our model
setting
for clustering, the model and prior can be expressed hierarchically as:
\begin{equation}
\label{eq:zip_mfm}
\begin{split}
	&y_{ij}\sim\text{ZIP}\left(\exp(\bm{x}_j^\top\bm{\beta}_{z_i}),
	\rho_{z_i}\right), \quad i=1, \ldots, n, \quad j=1,\ldots,J,\\
	&\bm{\beta}_h\sim \text{N}\left(\textbf{0},\Sigma_0\right),
	\qquad h=1,\ldots,k, \\
    &\rho_h\sim\text{U}\left(0,1\right),\qquad h=1,\ldots,k,\\
    &z_i \mid k, \pi \sim \sum_{h=1}^k  \pi_h \delta_h,\\
    &(\pi_1, \ldots, \pi_k) \mid k \sim \mbox{Dir} (\gamma,
	\ldots, \gamma),\\
	 &k \sim p(\cdot),    
\end{split}
\end{equation}
where  $p(\cdot)$ is a $\mbox{Poisson}(\psi)$ distribution truncated to be
positive through the rest of the paper, which has been proved by
\cite{miller2018mixture} and \cite{geng2019probabilistic} to guarantee consistency for the
mixing distribution and the number of groups, $\gamma=1$, $\delta_h$ denotes
the
Dirac measure, and $\Sigma_0$ is hyperparameter for base distribution of
$\bm{\beta}$'s. We refer to the hierarchical model above as MFM-ZIP.

\subsection{Theoretical Property}\label{sec:theory}

In this section, we study
the theoretical property of MFM-ZIP.
We assume that the parameter space~$\bm{\Theta^*}$ is the compact parameter
space for all the model parameters (i.e., mixture weights, regression
coefficients and zero inflated probability) given a fixed number of clusters.
The mixing measure is $G = \sum_{i=1}^k \pi_i \delta_{\theta_i}$, where
$\delta$
is the point mass measure, and $\theta_h = \{\bm{\beta}_h, \rho_h \}$ is the
collection of regression coefficients and zero inflation probability in cluster~$h$ for $h=1,\ldots,k$. 

Let $K_0$, $G_0$, $P_0$ be the true number of clusters, the true mixing
measure,
and the corresponding probability measure, respectively. Then the following
proposition establishes the posterior consistency and contraction rate for the
cluster number $K$ and mixing measure $G$. The proof is based on the general
results for Bayesian mixture models in \citet{guha2019posterior}.  
\begin{prop}\label{thm1} Let $\Pi_n(\cdot \mid \bm{y}_1,\ldots,\bm{y}_n)$ be
the
posterior distribution obtained from given a random sample
$\bm{y}_1,\ldots,\bm{y}_n$. Assume that the parameters of interest are
restricted to a compact space $\bm{\Theta^*}$. Then we have
\begin{align*} \Pi_n(K = K_0 \mid
\bm{y}_1,\ldots,\bm{y}_n) \rightarrow 1, ~\text{and}~~ \Pi_n (W(G,G_0)\lesssim
(\log n/n)^{-1/2} \mid \bm{y}_1,\ldots,\bm{y}_n) \rightarrow 1, 
\end{align*}
almost surely under $P_0$ as $n \rightarrow \infty$, where $W$ is Wasserstein
distance.
\end{prop}
Proposition~\ref{thm1} shows that our proposed Bayesian method is able to
correctly identify the unknown number of clusters and the latent clustering
structure with posterior probability tending to one as the number of
observations increases.

In order to prove the Proposition~\ref{thm1}, we need to verify the conditions
(P.1)--(P.4) in \cite{guha2019posterior} hold. Condition~(P.1) is satisfied
since we restrict our parameters of interest to a compact space~$\bm{\Theta^*}$
and uniform distribution and multivariate normal distribution are first-order
identifiable. Condition~(P.2) also holds since we assign an non-zero continuous
prior on $\bm{\beta}$'s and $\rho$'s on the parameters within a bounded
support.
Uniform distribution and multivariate normal distribution is sufficient for
Condition~(P.3). Condition~(P.4) holds since we choose a truncated Poisson
distribution on~$q(\cdot)$. The proof can be finished by using the results in
Theorem~3.1 of \cite{guha2019posterior}

\section{Bayesian Inference}\label{sec:bayesian_inference} For the hierarchical
ZIP model with MFM introduced in \eqref{eq:zip_mfm}, the set of parameters is
denoted as $\Theta =\{(\bm{\beta}_i,\rho_i, z_i, \bm{\pi},k): i = 1, \cdots,
n\}$. If we choose $(k-1) \sim \mbox{Poisson}(\psi)$ and $\gamma=1$ in
\eqref{eq:MFM}, the mixture weights~$\pi_1,\cdots,\pi_k$ can be constructed
following stick-breaking \citep{sethuraman1994constructive} approximation:

\begin{itemize}

	\item \textbf{Step 1.} Generate $\eta_1,\eta_2,\cdots \overset{\text{iid}}{\sim}
\text{Exp}(\psi)$,
	\item \textbf{Step 2.} $k=\min\{j:\sum_{k=1}^j \eta_k \geq 1\}$,
	\item \textbf{Step 3.} $\pi_h=\eta_h$, for $h=1,\cdots,k-1$,
	\item \textbf{Step 4.} $\pi_k=1-\sum_{h=1}^{k-1}\pi_h$.
\end{itemize}

Prior for the hyperparameter $\psi$ is $\text{Gamma}(1,1)$. With the prior
distributions specified above, the posterior distribution of these parameters
based on the data $D = \{\bm{y}_i,\bm{x}_j: i = 1,\cdots, n, j = 1, \cdots,
p\}$
is given by
\begin{align*}
	\pi(\Theta\mid D) &\propto L(\Theta\mid D) \pi(\Theta)\\
&= \prod_{i=1}^{n} f(\bm{y}_i,\bm{x}_1,\ldots,\bm{x}_j\mid
\bm{\beta}_{z_i},\rho_{z_i},z_i)\pi(\Theta),
\end{align*}
where $\pi(\Theta)$ is the joint prior for all the parameters. Due to the
unavailability of the analytical form for the posterior distribution
of~$\Theta$, we employ the MCMC sampling algorithm to sample from the posterior
distribution, and then obtain the posterior estimates of the unknown
parameters.
Computation is facilitated by the \textbf{nimble}\citep{de2017programming}
package in \textsf{R} \citep{Rlanguage2013}. The implementation code is given
in supplementary materials at
\url{https://ys-xue.github.io/MFM-ZIP-Basketball-Supplemental/}.
Another important task is to do the posterior inference for clustering labels.
We carry out posterior inference on the clustering labels based on Dahl's
method \citep{dahl2006model}, which proceeds as follows,

\begin{itemize}
\item \textbf{Step 1.} Define membership matrices $\mathcal{A}^{(t)}
=(\mathcal{A}^{(t)}(i,j))_{i,j \in \left\{1,\ldots,n\right\} } =
(\bm{1}(z_{i}^{(t)} = Z_{j}^{(t)}))_{n \times n}$, where $t = 1, \ldots, T$ is
the index for the retained MCMC draws after burn-in, and $\bm{1}(\cdot)$ is the
indicator function.
\item \textbf{Step 2.} Calculate the element-wise mean of the membership
matrices over MCMC draws $\bar{\mathcal{A}} = \frac{1}{T} \sum_{t=1}^{T}
\mathcal{A}^{(t)}$.
\item \textbf{Step 3.} Identify the most \textbf{representative} posterior
$\bar{\mathcal{A}}$ draw based on minimizing  
the element-wise Euclidean distance $\sum_{i=1}^{n} \sum_{j=1}^{n}
(\mathcal{A}^{(t)}(i,j) - \bar{\mathcal{A}}(i,j))^{2}$ among the retained $t =
1,\ldots,T$ posterior draws.
\end{itemize}
The posterior estimates of cluster memberships $z_1,\ldots,z_n$ and other model
parameters $\bm{\beta}$'s and $\rho$'s can be also obtained using Dahl's method
accordingly.

\section{Simulation Study}\label{sec:simulation}

\subsection{Simulation Setup}

We have two scenarios in our simulation, balanced type and imbalanced type. A
total of~75 players are separated to three different groups for each type.
Under
the balanced design, each group contains equal number of players. Under the
imbalanced design, the group sizes are~10, 35 and~30, respectively. The spatial
domain is the same as for the motivating data in
Section~\ref{sec:motivatedata}.
We generate data~$\{y_{i,j};~\forall i=1,\ldots,75; j=1,\ldots,1175\}$ from ZIP
model with different mean parameter and probability parameter of extra zeros.
In
our simulation setting, our covariates contains an intercept term and five
basis
function terms (see Section~\ref{sec:NMF}). Different values of coefficient
$\bm{\beta}$ are used: $\left(-1, 1.2, 0.95, 1.1, 1, 0.8\right)^\top$,
$\left(-0.4, 0.6, 0.7, 0.5, 0.8, 0.3\right)^\top$, and $\left(-0.9, 0.2, 0.1,
0.3, 0.2, 0.4\right)^\top$, corresponding to each cluster respectively. We set
the true probability parameter of extra zeros for each cluster to
be~$\left(0.1,
0.3, 0.4\right)$.

We examine both the estimation for number of groups as well as congruence of
group belongings with the true setting in terms of modulo labeling by Rand
index
\citep[RI;][]{rand1971objective}, the computation of which is facilitated by
the
\textsf{R}-package \textbf{fossil} \citep{vavrek2011fossil}. The RI ranges from
0 to 1 with a higher value indicating better agreement between a grouping
scheme
and the true setting. In particular, a value of 1 indicates perfect agreement.

\subsection{Simulation Results}

We run our algorithm with 7,000 MCMC iterations, with the first~2,000
iterations
as burn-in for each replicate. The chain length has been examined to ensure
convergence and stabilization. Proceeding to~100 separate replicates of data,
our
proposed algorithm was run, and~100 RI values are obtained by comparing with
the
true setting. We calculate cover rate for both scenario. For each scenario, we
also calculate the cover rate, which equals the percentage of replicates that
our proposed algorithm accurately recovers the true number of clusters. The
cover rate for each scenario are~98\% and~93\%, respectively. We also compare
our method to $K$-means algorithm, high dimensional supervised classification
and clustering \citep[hdc;][\textsf{R}-package
\textbf{HDclassif}]{berge2012hdclassif} and mean shift grouping. The mean shift
algorithm is a steepest ascent classification algorithm, where classification
is
performed by fixed point iteration to a local maxima of a kernel density
estimate. This method is originally from \cite{fukunaga1975estimation}, and an
implementation in~\textsf{R} can be found in the \textbf{meanShiftR} package
\citep{lisic2018}. Grouping recovery performances of all four methods are
measured using the RI. As $K$-means and the mean shift algorithm cannot infer
the number of clusters, such values need to be pre-specified, and we supply
them
with the number of clusters inferred by our method in each replicate. The
clustering performances are compared in Table~\ref{tab:comparison}. In both
designs, our proposed method have the highest RI, indicating its high accuracy
in clustering. $K$-means and hdc have RI greater than 0.9, but the average
performance is not as good as the proposed model. The mean shift algorithm,
however, yields the worst performance.

We provide parameter estimation in Table~\ref{tab:estimation}. For each of the
three $\bm{\beta}$'s, the average parameter estimate denoted by
$\Bar{\hat{\beta}}_{\ell,m}$ $\left(\ell = 1,\ldots ,75; m = 1,\ldots
,6\right)$
in 100 simulations is calculated as
\begin{align*}
    &\Bar{\hat{\beta}}_{\ell,m}=\frac{1}{100}\sum_{r=1}^{100}
    \hat{\beta}_{\ell,m,r}
\end{align*}
where $\hat{\bm{\beta}}_{\ell,m,r}$ denotes the posterior estimate for
the~$m$-th coefficient of player~$\ell$ in the $r$-th replicate. We use
different metrics to evaluate the posterior performance. Those metrics
including
the mean absolute bias (MAB), the mean standard deviation (MSD), the mean of
mean squared error (MMSE) and mean coverage rate (MCR) of the 95\% highest
posterior density (HPD) intervals in the following ways:
\begin{align*}
	\text{MAB}&=\frac{1}{75}\sum_{\ell=1}^{75}\frac{1}{100}\sum_{r=1}^{100}
	\abs{\hat{\beta}_{\ell,m,r}-\beta_{\ell,m}},\\
	\text{MSD}&=\frac{1}{75}\sum_{\ell=1}^{75}\sqrt{\frac{1}{99}\sum_{r=1}^{100}
	\left(\hat{\beta}_{\ell,m,r}-\Bar{\hat{\beta}}_{\ell,m}\right)^2},\\
	\text{MMSE}&=\frac{1}{75}\sum_{\ell=1}^{75}\frac{1}{100}\sum_{r=1}^{100}
	\left(\hat{\beta}_{\ell,m,r}-\beta_{\ell,m}\right)^2,\\
	\text{MCR}&=\frac{1}{75}\sum_{\ell=1}^{75}\frac{1}{100}\sum_{r=1}^{100}
	1_{\left\{\hat{\beta}_{\ell,m,r}\in \text{95\% HPD interval}\right\}},
\end{align*}
where $1_{\{\}}$ denotes the indicator function. The four metrics for
each~$\beta$ under the balanced and imbalanced designs are presented in
Table~\ref{tab:estimation}. With high clustering accuracy as indicated by the
RI, the estimated~$\bm{\beta}$ for each cluster is close to its corresponding
true value, which can be reflected by the small numerical values in the MAB
column. The estimation performance is stable, in the sense that the MSD values
are also small. The MCR under the balanced design fluctuate around its nominal
value of~0.95, and under the imbalanced design, the values are overall lower
due to the influence of mis-clustered players, but still remain close to or
greater than~0.9.

\begin{table}
\caption{\label{tab:comparison} Comparison of clustering performance for the
proposed method and three other competing approaches.}
\centering 
	\begin{tabular}{lccccc}
\hline
Type & Cover Rate & $\text{RI}_{\mbox{mfm}}$ & $\text{RI}_{\mbox{kmeans}}$ &
		$\text{RI}_{\mbox{hdc}}$ & $\text{RI}_{\mbox{meanshift}}$ \\
\hline
		Balanced & 98\% &0.9955 &0.9364  & 0.9344&0.6757\\
		Imbalanced & 93\% &0.9836  &0.9581&  0.9735 &0.6126\\
\hline
	\end{tabular}
\end{table}

\begin{table}
\caption{\label{tab:estimation} Performance of parameter estimates
under the two true cluster designs.}
\centering
	\begin{tabular}{lcccccc}
\hline
Type & 
 &MAB &MSD &MMSE &MCR \\
\hline
Balanced & $\beta_{0}$ & 0.01610 &  0.0421 &  0.001930 &  0.933 \\
		& $\beta_{1}$ & 0.01340 &  0.0322 &  0.001060 &  0.927 \\
		& $\beta_{2}$ & 0.01190 &  0.0279 &  0.000884 &  0.943 \\
		& $\beta_{3}$ & 0.00928 &  0.0272 &  0.000788 &  0.960 \\
		& $\beta_{4}$ & 0.01150 &  0.0266 &  0.000904 &  0.930 \\
		& $\beta_{5}$ & 0.00920 &  0.0235 &  0.000604 &  0.920 \\
\hline
Imbalanced & $\beta_{0}$ & 0.02680 &  0.0663 &  0.005950 &  0.901 \\
    	  & $\beta_{1}$ & 0.02340 &  0.0521 &  0.003160 &  0.911 \\
    	  & $\beta_{2}$ & 0.02030 &  0.0504 &  0.002790 &  0.896 \\
    	  & $\beta_{3}$ & 0.01880 &  0.0424 &  0.002360 &  0.921 \\
    	  & $\beta_{4}$ & 0.01960 &  0.0491 &  0.002790 &  0.916 \\
    	  & $\beta_{5}$ & 0.01740 &  0.0380 &  0.001770 &  0.912 \\
\hline
	\end{tabular}
\end{table}

\section{Real Data Application}\label{sec:realdata}

In this section, we apply
the proposed method to the analysis of players' shot data in the 2017-2018 NBA
regular season. Only the locations of shots are considered regardless of the
player's positions on the court (e.g., point guard, power forward, etc.). We
run
15,000 MCMC iterations and the first 5,000 iterations as burn-in period. The
result from the MFM-ZIP model suggests that the 191 players are to be
classified
into four groups. The sizes of the four groups are 29, 110, 48 and 4
respectively. We visualize the shot attempt counts made by four selected
players
on blocks of the court in Figure~\ref{fig:realdata}. The players for each group
are shown in Section~3 of the
supplementary materials.

\begin{figure}
    \includegraphics[width=\textwidth]{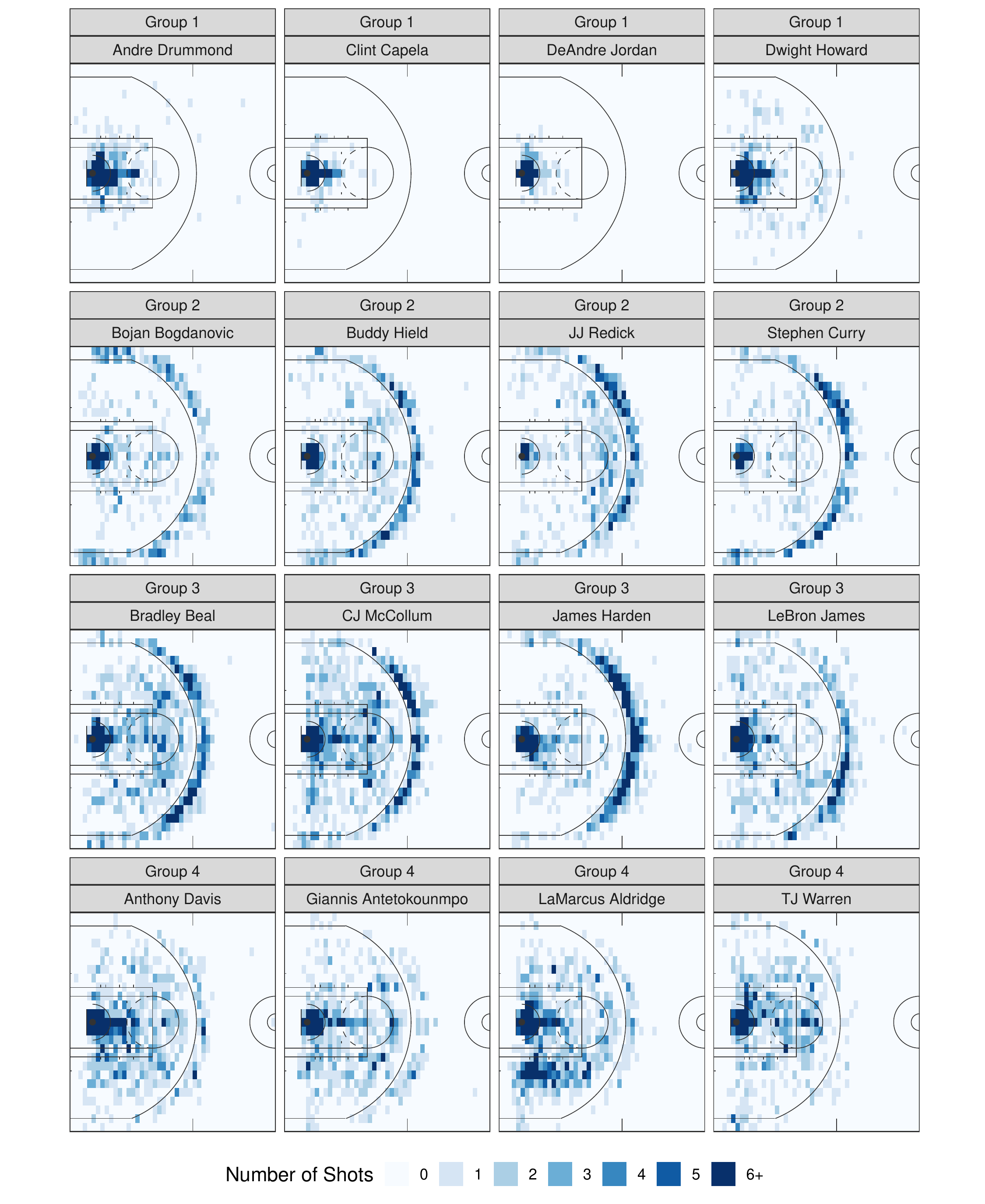}
    \caption{\label{fig:realdata} Visualization of shooting patterns for four
selected players from each group.}
\end{figure}

\begin{table}
\caption{\label{tab:realdataestimation} Performance of parameter estimates
under the real data.}
\centering
	\begin{tabular}{cccccccc}
\hline
& $\rho$ &$\beta_{0}$ &$\beta_{1}$ &$\beta_{2}$
&$\beta_{3}$ &$\beta_{4}$ &$\beta_{5}$ \\
\hline
		Group 1 & 0.583  & -1.857 & 0.256 & 0.573 & 1.374 & 0.531 & 1.061\\
		Group 2 & 0.618  & -0.409 & 0.731 & 0.200 & 0.359 & 0.581 & 0.392\\
		Group 3 & 0.515  & -0.147 & 0.511 & 0.237 & 0.505 & 0.493 & 0.474\\
		Group 4 & 0.322  & -1.205 & 0.258 & 1.101 & 1.056 & 0.426 & 0.766\\
\hline
	\end{tabular}
\end{table}

Several interesting observations can be made from Figure~\ref{fig:realdata}. It
can be seen that each group of players has their own favorite shooting
locations. Players in Group~1 make the most shots neat the hoop, which is
confirmed by the regression coefficients in Table~\ref{tab:realdataestimation}
as their coefficients for the third and fifth basis functions are the largest.
Clint Capela and DeAndre Jordan, for example, are both good at making
alley-oops
and slam dunks. Andre Drummond and Dwight Howard are also centers who rarely
leave the painted area.

Players in Group~2 make the most shots beyond the three-point line, as they
have
the largest parameter estimates for the first and fourth basis function when
compared to other groups. As shown in Figure~\ref{fig:realdata}, JJ Redick and
Stephen Curry are both well-known shooters.

A first look at the plots for Group~3 indicates that players in this group are
able to make all types of shots, including three-pointers, perimeter shot, and
also shots over the painted area. We find the players in this groups are often
the leaders in their teams, and usually have the most posession. The parameter
estimates also confirm the observation. Their $\hat{\beta}_0$ is the largest
among all groups, indicating an overall higher probability for making shots.
Compared with Group~2, their shots are more evenly distributed, which can be
reflected by the larger parameter estimate for the basis functions
corresponding
to areas within the three-point line.

For Group~4, we find that most of their shots are close to the hoop and around
the perimeter, and they have fewer shot around or beyond the three-point line.
From the estimation result, the coefficient for the second and third basis
functions are larger than other basis functions, and similar in value. In
addition, their~$\hat{\beta}_1$ is the second smallest among all four groups
while~$\hat{\beta}_4$ is the smallest, which again indicates their disfavor of
shooting beyond the three-point line. Note that the presented analysis is based
on 2017--2018 regular season, which was before Giannis Antetokoumpo increased
his three-point shots in the 2019--2020 season.

\section{Discussion}\label{sec:discussion} 

Based on theoretical justifications and empirical studies, our proposed methods
successfully solve the three challenges raised in Section~\ref{sec:intro}.
Based
on the results shown in Section~\ref{sec:simulation}, our proposed methods
accurately estimates the parameters in the ZIP model and recovers the number of
clusters and clustering configurations with different proportions of zero
counts. Compared with several benchmark clustering methods such as $K$-means,
high dimensional supervised classification and clustering, and mean shift
grouping, our methods have higher clustering concordance without any tuning
steps.

In the analysis of the NBA shot charts data, four field goal attempt patterns,
their corresponding zero inflation probabilities, and regression coefficients
of
each group are identified. The results provide valuable insights to players,
coaches, and managers. The players can obtain more descriptive analysis of
their
current offense patterns, and hence develop customized training plans with
pertinence; the coaches can organize their offense and defense strategies more
efficiently for different opponents; the mangers will make better data-informed
decisions on player recruiting and trading during offseason.

There are several possible directions for further investigation. Spatial
correlation over the court is accounted for nonparametrically by the basis
functions.
Considering
either stationary or nonstationary model-based
spatial correlation in our proposed modeling framework is
a natural extension. In this paper, our posterior sampling is based on
stick-breaking representations. Developments of more scalable inference
algorithms (e.g., variational inference) are critical for large scale data.
Finally, building a heterogeneity learning model with auxiliary information
from
different players, such as age and position on the court,
 merits future research from both methodological and applied
perspectives.

\end{document}